# The Decision-Theoretic Interactive Video Advisor

Hien Nguyen*, Peter Haddawy*†

Decision Systems and Artificial Intelligence Lab*
Dept. of EE &CS
University of Wisconsin-Milwaukee
Milwaukee, WI 53201
{nguyen,haddawy}@cs.uwm.edu

Intelligent Systems Lab†
Faculty of Science & Technology
Assumption University
Bangkok 10240, Thailand
haddawy@isl.s-t.au.ac.th

**Abstract**
The need to help people choose among large numbers of items and to filter through large amounts of information has led to a flood of research in construction of personal recommendation agents. One of the central issues in constructing such agents is the representation and elicitation of user preferences or interests. This topic has long been studied in Decision Theory, but surprisingly little work in the area of recommender systems has made use of formal decision-theoretic techniques. This paper describes DIVA, a decision-theoretic agent for recommending movies that contains a number of novel features. DIVA represents user preferences using pairwise comparisons among items, rather than numeric ratings. It uses a novel similarity measure based on the concept of the probability of conflict between two orderings of items. The system has a rich representation of preference, distinguishing between a user's general taste in movies and his immediate interests. It takes an incremental approach to preference elicitation in which the user can provide feedback if not satisfied with the recommendation list. We empirically evaluate the performance of the system using the EachMovie collaborative filtering database.

## 1 INTRODUCTION

Content-based and collaborative filtering have become popular approaches for eliciting user preferences in order to recommend items of interest[1]. Representation and elicitation of preferences have long been studied in Decision Theory (Keeyney & Raiffa 1976), but surprisingly almost no work in the area of collaborative filtering systems has made use of the wealth of techniques and formalism available. This paper describes the Decision-Theoretic Interactive Video Advisor[2] (DIVA 1.0), a collaborative filtering system that provides movie recommendations. The guiding principle behind the system design is the accurate representation and efficient elicitation of user preferences, following decision-theoretic principles wherever they apply. Following this design methodology has led to a number of novel features that provide DIVA with distinct advantages over other collaborative filtering systems.

All collaborative filtering systems to date represent user preferences with numerical ratings. In contrast, DIVA uses the standard decision-theoretic notion of a preference order, i.e., pairwise comparisons among movies. This provides a fine-grained representation of preference without forcing the user to think about a large number of rating categories.

Most collaborative filtering systems, e.g. (Hill et al. 1995, Shardanand & Maes 1995), determine similarity between preferences of different users following the technique used in GroupLens (Resnick 1994), which is based on the Pearson correlation coefficient. In contrast, DIVA uses a measure based on the concept of the probability of conflict between pairwise rankings. We show that our measure has several practical advantages over the GroupLens measure. We empirically demonstrate that use of our similarity measure results in more accurate recommendations than use of the GroupLens measure.

While people can be characterized as having a general taste in movies, what they are interested in seeing on any one occasion may deviate from this. DIVA supports this dynamic notion of preference by distinguishing between long- and short-term preferences.

In order to minimize the amount of time the user needs to spend providing preference information to the system, DIVA supports the notion of incremental elicitation of preferences by permitting the user to provide feedback if he is not satisfied with the list movies DIVA recommends. This feedback takes into account the distinction between long- and short-term preferences.

---
[1] AAAI Workshop on Recommender Systems, Madison, July 1998. AAAI Technical Report WS-98-08.
[2] Available at www.uwm.edu/~dsail or isl.s-t.au.ac.th/

The rest of this paper is organized as follows: section 2 provides an overview of DIVA's functionality. Section 3 describes the overall architecture and key algorithms. Section 4 presents the results from our empirical evaluation of the system. Section 5 discusses related work, and section 6 presents conclusions and directions for future research.

## 2 OVERVIEW OF DIVA

DIVA's design emphasizes ease and accuracy of preference elicitation. In order not to overburden the user with unnecessary questions, DIVA takes an incremental approach to preference elicitation. It starts by eliciting some preference information and then quickly provides a list of recommendations. If the user is not satisfied with the recommendations, the user can critique them in order to provide additional preference information.

To accurately represent user preferences, we attempt to account for the dynamic nature of preferences. While people have general taste in movies, the type of movie they would like to see will typically vary from one occasion to another, based on many environmental factors, such as what movies they recently saw. Thus we separate the elicitation of long- and short-term preferences and we combine the two types of preference in a natural and intuitive fashion.

### 2.1 REGISTERING NEW USERS

A new user to DIVA is asked to provide a user login name and a password, which are used to index his long-term preference information whenever he accesses the system. The user is then shown an alphabetical list of all movies in the system and asked to indicate some he particularly liked, disliked, and some he thought were about average (Figure 1). Experience has shown that a user must classify a minimum about 5 movies in each category to get reasonable recommendations from the system. These sources of information are used to build an initial preference structure for the user. Figure 1 shows the preferences specified by user *testaccount* who likes comedies (*Bob Roberts, GroundHog Day, Bread and Chocolate*).

### 2.2 SEARCHING

After submitting his preferences, the user may request recommendations. This is done via the search page that allows the user to constrain the search, as shown in Figure 2. The user may specify actors and actresses, directors, genres, professional star ratings, countries of production, release years, MPAA ratings, and the running time that he is particularly interested in. These constraints give the user a way of indicating what kind of movie he is particularly interested in watching at the current time, i.e. his short-term preferences. They act as an initial filter on the entries in the movie database. The movies that satisfy the constraints are then ranked according to the user's preferences. The behavior we have endeavored to embody in DIVA is that of someone who knows you well acting on your behalf with some instruction from you. For example, suppose you ask a close friend to go to the video store and rent an adventure film that takes place in Africa. Within the context of that constraint, she would use her knowledge of your taste in movies, e.g. a preference for films with excellent cinematography, to select the film she thinks you would be most likely to enjoy.

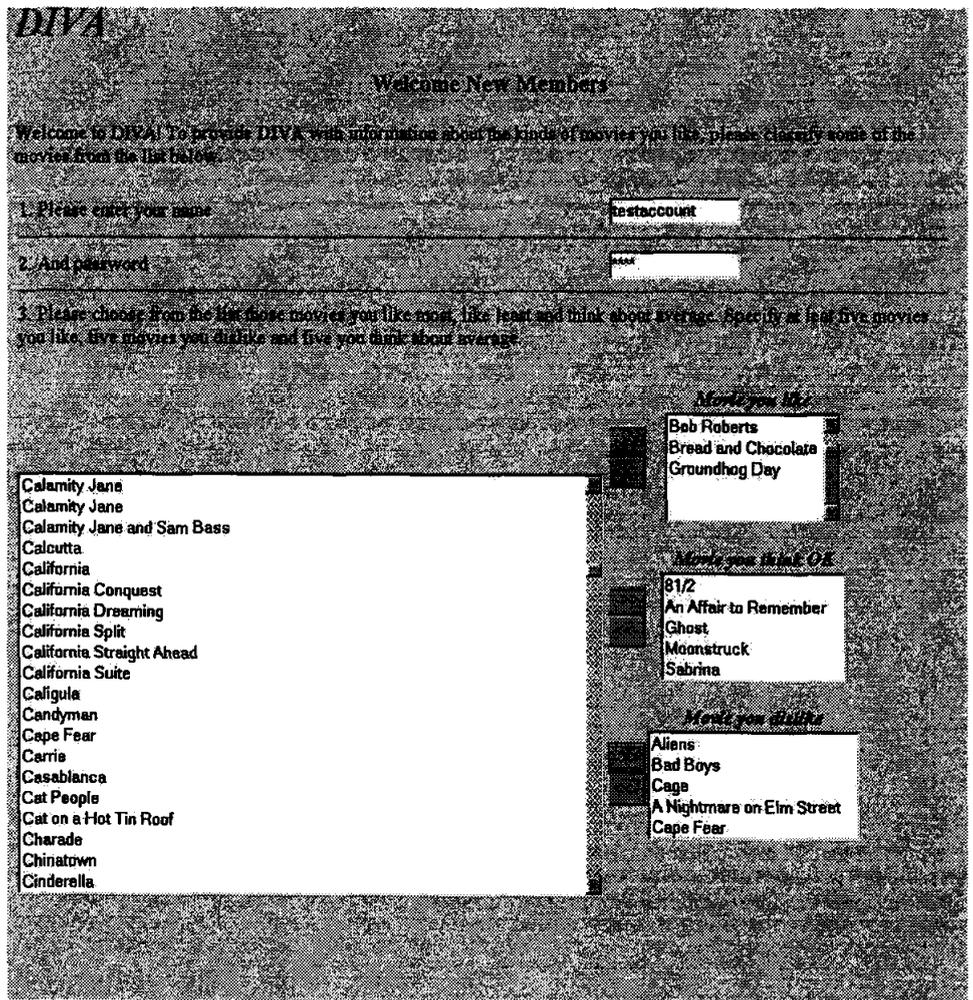

**Figure 1:** A screenshot of registration window.

In the example shown in Figure 2, although the user *testaccount* generally likes comedies, at the present time he is interested in searching for a crime film. The recommendation list is displayed on a separate page, as shown in Figure 3. In this example, the user's short-term preferences are blended with the user's long-term preferences nicely. The recommendation list contains only movies that belong to both genres *crime* and *comedy*. By clicking on any of the movie titles in the recommendation list, the user can see all the attribute information about that film.

### 2.3 PROVIDING FEEDBACK

If the user is not interested in any of the films in the recommendation list, he can provide feedback to the system and request another search. We distinguish between feedback concerning long-term (second feedback column) and short-term (first feedback column) preferences. If the user has seen some of the movies in the list, he can indicate whether he particularly liked or disliked the film. Even though he may not have seen the film, the user may know enough about it to be quite confident that he would not enjoy seeing it. This is all feedback concerning long-term preferences and is added to the stored user preference model. For movies the user has not seen, DIVA asks him to indicate if they are close to what he is interested in watching currently or if they are far from that. This is feedback concerning short-term preferences and is used only in the current search session. After providing feedback, the user can click the button "Continue Search" to obtain a new list of recommendations. Otherwise, he can start an entirely new search by clicking "New Search" (see Figure 3).

**Figure 2:** A screenshot of the search window.

## 3   ARCHITECTURE AND ALGORITHMS

We retrieve movies by first computing the preference ranking over all movies in our movie database. Then we remove any movies that do not satisfy the user's short-term constraints and display the top $n$ movies. If not that many movies satisfy all the constraints, we relax the constraints by disjoining them.

### 3.1   INITIAL PREFERENCE ELICITATION

A user's complete set of preferences in a domain with no uncertainty (such as the movie domain) can be represented as a total ordering over the items in the domain. A subset of the user's preferences then corresponds to a partial order over the items. An initial set of pairwise preferences among movies is obtained from the user's Like, OK, and Dislike lists (Figure 1). This gives us a partial order over movies: every movie in the Like list is considered preferred to every movie in the OK list, which is in turn preferred to every movie in the Dislike list (will be referred as *LDO partial order*). Figure 6(b) (without the dashed links and the filled node) shows the preferences obtained from the Like, OK, and Dislike lists of the user in Figure 1.

## 3.2 PREDICTING PREFERENCES

The directly elicited user preferences are augmented with preference information from a case base of user preference information, using the technique described in (Ha & Haddawy 1998). See that paper for details. The case base contains partial preference structures for all users of the system. We compute the similarity between the active user's initial partially elicited preference structure and the preference structures in the case base. We then use the preferences of the most similar stored structure to supplement the directly elicited preferences.

In order to find the closest matching preference structure, we need a method of computing the similarity between two preference structures. We define the *dis*-similarity between two complete preference structures as the probability that two randomly chosen elements are ranked differently by the two preference structures. This similarity measure satisfies all the properties of a distance metric and has range [0,1].

However, the preference information in our case base and the preference information for the active user are only partial. So we need a similarity measure over partially specified preference structures, i.e. partial orders. A partial order can be viewed as a set of total orders that are consistent with it, where the consistency means that any relation between any pair of elements that holds with respect to the partial order also holds with respect to the total order. The consistent total orders of a partial order are called *linear extensions*.

We define the *dis*-similarity between two partial preference structures simply as the average *dis*-similarity between all pairs of linear extensions in the two sets. This measure also has range [0,1] and satisfies the triangle inequality, but it is not a metric because the distance between two identical partial orders that are not complete orders is always positive. Fortunately, this is desirable if the two orders represent the preferences of two different users, since the complete preference structures of the two may actually differ.

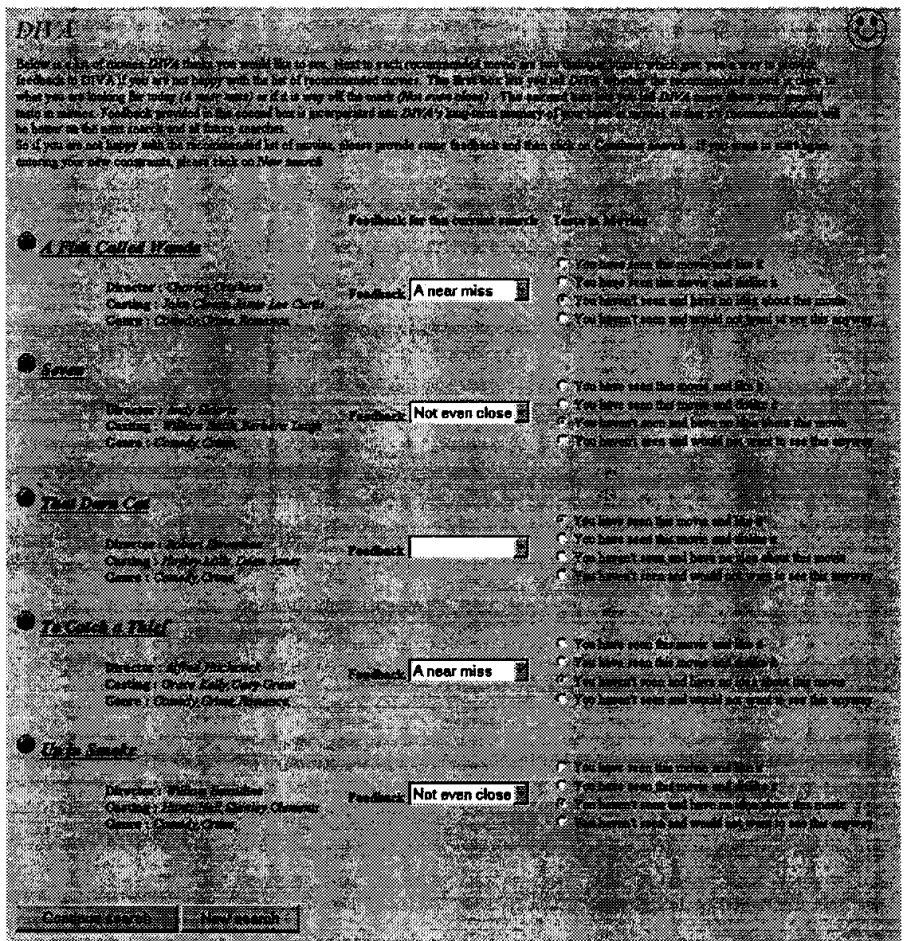

**Figure 3:** The recommendation list of the search and the feedback.

Computing this distance measure is closely related to the problem of computing the number of linear extensions of a finite partial order. That problem is known to be #P – complete (Brightwell & Winkler 1991). We use the Markov chain based approximation algorithm of (Bubley & Dyer 1998), which almost uniformly samples the space of linear extensions and runs in polynomial time. To generate each sample linear extension, the algorithm involves running a Monte Carlo simulation of a Markov chain for a fixed number of iterations. After computing the distance between the active user and each preference structure in the case base, we choose the most similar preference structure and then choose the sampled linear extension for the active user that is most similar to that preference structure. In this way, we retain all the directly elicited user preferences and obtain a complete preference structure for the active user, guided by the preferences in the case base. In effect, the preference structures in the case base are used as attractors, indicating in which direction to complete the partial preference structure of the active user. In order to save computation time, we compute the similarity only over the top 100 ranked movies in each pair of partial preference structures. The figure 4 illustrates the main idea of the algorithm and the figure 5 shows the pseudo code of the algorithm to compute the preference ranking for the active user.

### 3.3 INCORPORATING USER FEEDBACK

DIVA permits the user to provide feedback concerning long-term and short-term preferences. The long-term preference feedback is added to the stored partial preference structure obtained from the original "Like","OK", "Dislike" lists. If the user saw a movie and liked it, the movie is added to the Like list. It is shown in Figure 6(b) as a filled circle associated with dashed lines. If the user saw a movie and disliked it or if he didn't see a movie but is confident that he wouldn't like it, the movie is added to the Dislike list. The short-term preference feedback is stored separately from the long-term preferences and is forgotten after the current search is ended. Movies the user tags as being *near misses* are considered preferred to those tagged as being *not even close*. A preference link is created from every *near miss* movie to every *not even close* movie, as shown in Figure 6(a). The preferences obtained from the user's feedback are added to the LDO partial order, which is directly derived from the like, dislike, and ok lists. The resulting partial preference structure is used as input to the *PreferenceRanking* algorithm and a new recommendation list is generated.

### 3.4 DATABASES

DIVA contains both a movie database and a user preference case base. We built the initial case base using the EachMovie collaborative filtering database provided by Digital Equipment Corporation. The Digital systems research center assembled the database by running a collaborative filtering system for 18 months. The EachMovie database contains 72,916 users, 2,811,983 numeric ratings and 1628 movies. To represent user preferences, EachMovie uses a numeric rating scale ranging from 0 (awful) to 1 (excellent), in increments of 0.2. To populate our case base, we chose 500 users out of the 72,916 users such that each user rated at least 20 movies. The average number of ratings per user over all 500 users is around 70 movies.

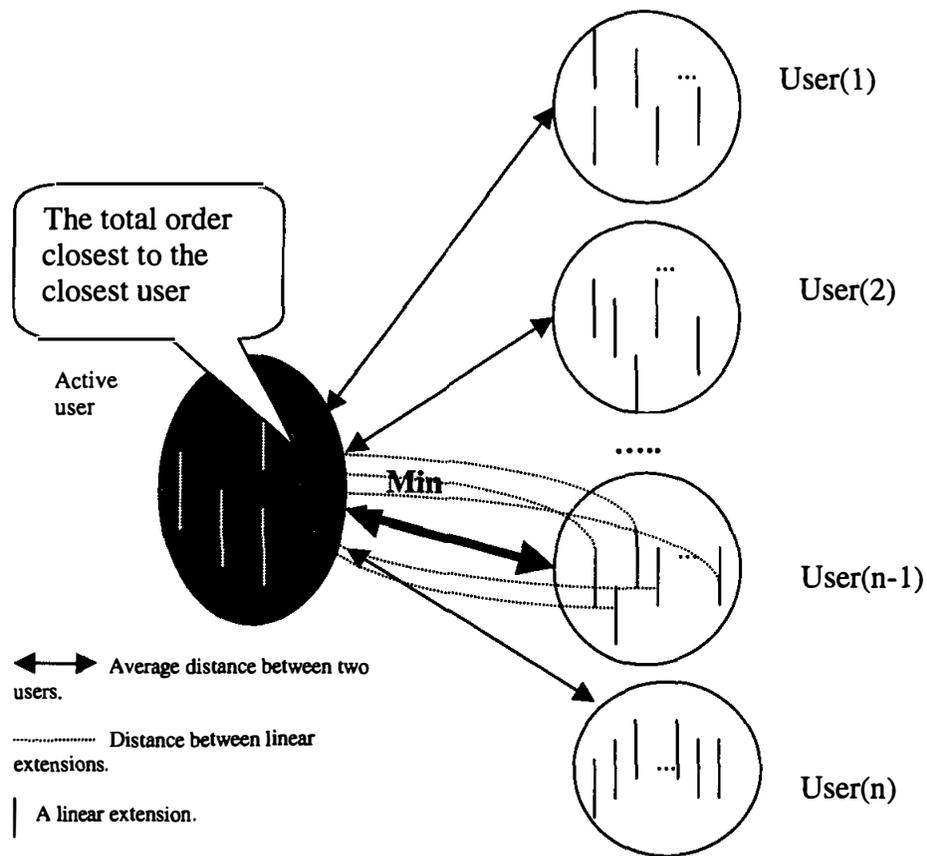

**Figure 4**: An illustration of algorithm to compute the preference ranking.

Since the EachMovie database did not contain sufficient attribute information for our purposes, we obtained movie attributes from the Microsoft Cinemania 94 CDROM. Our movie database contains 2000 movies with 9 attributes including actors/actresses, director, genre, professional film critic star rating, MPAA rating, country of production, running time, year released and title. Of the 1628 movies in EachMovie, only 500 appear in the Cinemania database. So we removed from the EachMovie rankings for all those movies that are not present in Cinemania. We then converted the preference structures represented as numeric ratings in EachMovie to partial order preference structures in the obvious way: For each user, a movie $s_1$ is preferable to movie $s_2$ if the user's rating for movie $s_1$ is greater than that for movie $s_2$.

## 4    EMPIRICAL ANALYSIS

We evaluated the quality of DIVA's recommendations and those generated using the GroupLens (Resnick 1994) collaborative filtering technique over our database of 500 users. We used the precision and recall metrics commonly used in information retrieval (Salton 1983) as our evaluation criteria. In the present context, precision indicates how many movies in a recommended list were actually liked by the user. Recall indicates how many movies out of all movies liked by the user were predicted correctly. We formed our test set by randomly taking out 10 users from the case base. For each user in the test set, we divided the partial preference structure into an observed set $O_1$ and unobserved set $O_2$. The observed set contained the OK and Dislike lists and three movies from the Like list. We ran the evaluation of DIVA by using the preference structure of the observed set to simulate elicitation from a new user and predicted unobserved items in $O_2$, i.e. the bulk of the Like list. We generated a recommendation list of length 1/6 of all movies that the user rated, which for all users in the experimental set was a subset of the number of items in $O_2$. We chose the value 1/6 in order to maximize precision at the expense of recall because from a user standpoint the desirability of the items in the recommendation list is more important than finding all items of interest in the data base.

```
PreferenceRanking( U1 // the active user's partial preference structure
                   U2[ ] // array of users' preference structures in the case base)
{
    // Sample a number of linear extensions for the active user.
    LI1 <- LinearExtensionInit(U1)
    For (number of linear extensions)
        LI_ij <- Sampling (LI_1, U1, stepSampling)

    // Compute the distance measure and find the closest matching user.
    For (U2i = preference structure of each user in the case base) {
        TotalDistance <- 0; Min_j <- -1
        For (number of linear extensions) {
            LI_2 <- LinearExtensionInit(U2)
            For (number of linear extensions) {
                LI_2j <- Sampling(LI_2, stepSamplings)
                Avg <- Avg + ComputeDistance(LI_ij, LI_2j)
            } // end of for
            Avg <- Avg / number of linear extensions
            If (the first sampled linear extension of the active user OR
                Min_j > Avg) {
                    Min_j <- Avg; LI_min <- LI_ij
            }
            TotalDistance <- TotalDistance + Avg
        } // end of for
        TotalDistance <- TotalDistance / number of linear extensions
        If (the first user in the case base OR
            MinDistance > TotalDistance) {
                MinDistance <- TotalDistance; LI_result <- LI_min
        }} // end of for
    return(LI_result)
}
```

**Figure 5:** Pseudo code of the algorithm to compute the preference ranking.

We ran 100 experiments with each of 10 users. We set the number of iterations in our sampling algorithm to 50, 100, and 150 and the number of linear extensions to 10, 30, and 50. The results are shown in Table 1, averaged over all 10 users. Note that if the number of linear extensions is greater than 30 and the number of iterations is greater than 100, the precision and recall do not change much.

We ran the same set of experiments using the GroupLens collaborative filtering technique. GroupLens (Resnick et al 1994, Konstan et al 1998) is a collaborative filtering system that helps News readers to find articles of interest. Each News reader rates a number of articles on a 5 point numeric scale. The system uses these ratings to determine which users are most similar to each other and then predicts how much the user will like new articles based on ratings from similar users. GroupLens uses a similarity metric based on the Pearson correlation coefficient. The measure assumes that preferences are represented with numeric ratings. Thus, for this portion of the experiment we worked with the original numeric rating representation of the preferences in our case base of 500 users. The similarity between two users is defined as the Pearson correlation coefficient over the intersection of the sets of movies that each rated. Predictions for movies that the active user did not rate are then produced by using the correlation to compute a weighted sum over all users in the case base. If a user is highly correlated with the active user, the weighting scheme assigns a large weight to that user's votes. If a user is highly negatively correlated, the weighting scheme assigns a large weight to the opposite of that user's votes. The precision and recall using the GroupLens technique were 65% and 35% respectively. The best precision and recall for DIVA were 86% and 40%, respectively, so the case-based elicitation technique in DIVA outperformed the GroupLens technique along both dimensions.

This result is not surprising given the properties of the similarity measures. The GroupLens measure has the disadvantage that it is insensitive to the size of the set of movies that both users have rated. In an extreme case, the preferences of two users can be maximally similar if they have only one movie in common that they rated and they agree on that rating. In contrast, the distance measure used in DIVA considers all movies that the two users have rated. Experiments we conducted to examine the set of users ranked similar to a given user showed that the DIVA measure does not rate two users as being similar unless their preferences agree over a set of reasonable size. Because it measures similarity only over the intersection of movies that two users have rated, the GroupLens measure does not satisfy the triangle inequality.

In addition to being intuitively desirable, the triangle inequality can be exploited to reduce computational effort by computing bounds on similarity. Note that there is no obvious way to apply the Pearson correlation coefficient to the union of the set of movies that two users have rated.

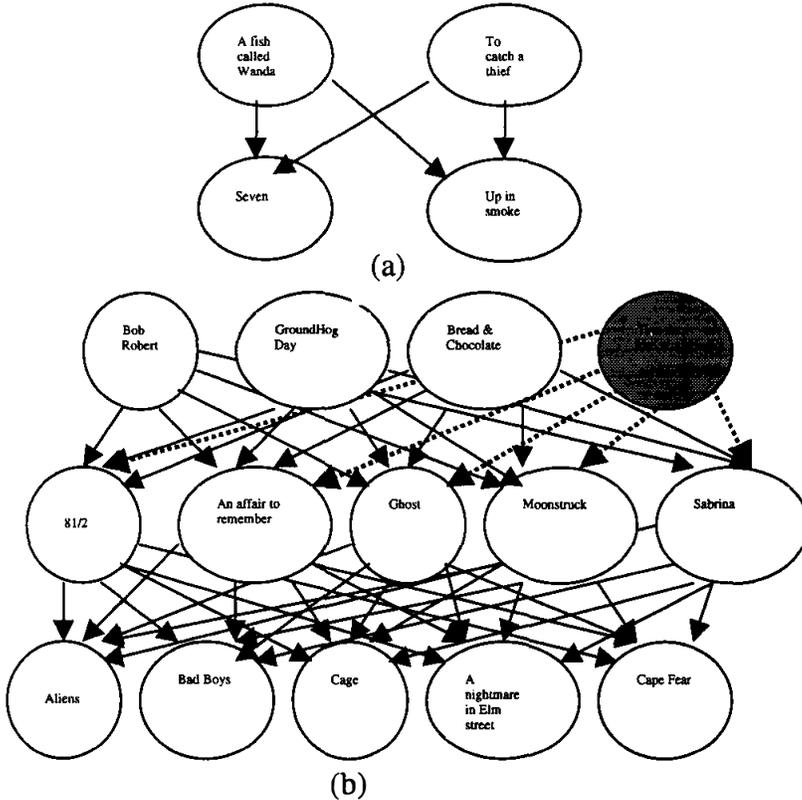

Figure 6: (a) Preference structure of the user's short-term feedback (b) Preference structure combining long-term feedback with initial *LDO partial order*.

| Number of linear extensions | 10 | 30 | 50 |
|---|---|---|---|
| Number of iterations | 50,100,150 | 50,100,150 | 50,100,150 |
| Precision | 80%,83%,83% | 81%, 86%,86% | 84%,85%,86% |
| Recall | 38%,40%,40% | 38%,40%,40% | 40%,40%,40% |

Table 1: Average precision and recall for DIVA's case-based elicitation algorithm.

## 5   RELATED WORK

The Automated Travel Assistant (ATA) (Linden 1997) is a system for recommending airline flights. It uses decision-theoretic techniques to determine user preferences and takes an incremental approach to elicitation. The ATA system elicits some user preferences concerning airline, price, and non-stop *vs* indirect, and combines these with default preferences (prefer cheaper flights, prefer fewer stops) to obtain a complete preference structure. It then presents the top ranked flights to the user, as well as some extreme solutions: cheapest flight, most direct flight. If the user is not satisfied with the recommendations, he can modify the model of his preferences by manipulating a graphical representation of the function used to represent the preferences. ATA and DIVA differ in their approach to incremental elicitation. While DIVA incrementally obtains increasing amounts of preference information from the user, ATA allows the user to incrementally modify an always complete preference structure. The user provides feedback to ATA by directly manipulating the system's internal representation of preferences. In contrast, the feedback mechanism in DIVA is intended to allow the user to communicate about the proposed solutions in a way that is natural for the application domain. Whereas ATA applies a single set of defaults to every user, the defaults used in DIVA are determined by searching the case base, using the initially elicited preferences. ATA has the advantage that its approach can be applied to domains with uncertainty because it can be used with utility functions. We are currently working on a similarity measure for utility functions.

Basu et al (Basu 1998) describes an approach to generate recommendations that combines content information with collaborative information. The content information includes values for 26 features in the movie selection domain such as genre, actors/actresses, and director. The case base contains 260 users and 45,000 movie ratings on scale of 1 – 10. The average precision and the recall of the system are 83% and 34%, respectively. Our precision and recall as seen in Table 1 are 86% and 40% (for 30 linear extensions and 100 iterations of the sampling algorithm). But this is only a rough comparison of the two approaches since the two systems were evaluated on different collaborative case bases. DIVA incrementally obtains user preferences through feedback, while Basu et al's approach involved no interaction with the user.

## 6   CONCLUSIONS AND FUTURE RESEARCH

This paper has described the first attempt to use decision-theoretic techniques in design of a collaborative filtering system. DIVA 1.0 addresses several difficult problems in building recommender systems, including the distinction between a user's general tastes and his immediate interests, and provision for the user to provide feedback if unsatisfied with the recommendations. Several difficult technical problems remain to be solved. Our representation of short-term preferences is rather crude. We would like to be able to represent short-term preferences using as rich a representation as that for long-term preferences and to be able to merge the two even when conflicts exist. In addition to distinguishing between long and short-term preferences, a recommender system should be able to account for the fact that people's preferences evolve over time. This would require a system to keep track of the time that each piece of preference information was obtained and to notice when newly expressed preferences conflict with older preferences.

A problem for DIVA and other collaborative filtering systems is the computational cost of computing similarity when the case base becomes very large. We are examining the user of hierarchical clustering of the case base in order to reduce the computational cost of case retrieval. An interesting question is how much accuracy we lose in the recommendations produced *vs* how much computation time we save.

To save computation time, we currently use the rough heuristic of computing similarity over only the top 100 ranked movies. We would like to experiment with other heuristics that involve sampling the preference structure in other ways.

The elicitation of pariwise preferences using the Like, OK, and Dislike lists results in a ranking of movies into only three categories. But the internal representation supports an arbitrary number of categories. We are currently working on augmenting the interface so that the user can graphically specify a richer set of preferences by directly positioning movies within each list on a vertical axis. We expect this will improve the accuracy of the recommendations.

## Acknowledgements

This work was partially supported by NSF grant IRI-9509165. The EachMovie database was generously provided by Digital Equipment Corporation. Thanks to Preerapol Meomeeg for help with the implementation and to Vu Ha, Hanh Pham and Thuong Doan for useful discussions.